\documentclass{article} 
\usepackage{iclr2020_conference,times}
\usepackage{xspace}

\usepackage{amsmath,amsfonts,bm}

\def\eqref#1{equation~\ref{#1}}

\def\1{\bm{1}}

\def\rc{{\textnormal{c}}}

\def\rj{{\textnormal{j}}}

\def\rx{{\textnormal{x}}}

\def\va{{\bm{a}}}

\def\mI{{\bm{I}}}

\def\mX{{\bm{X}}}
\def\mY{{\bm{Y}}}

\DeclareMathAlphabet{\mathsfit}{\encodingdefault}{\sfdefault}{m}{sl}
\SetMathAlphabet{\mathsfit}{bold}{\encodingdefault}{\sfdefault}{bx}{n}

\newcommand{\R}{\mathbb{R}}

\newcommand{\acronym}{GAN-TTS\xspace}
\newcommand{\rwd}{$\mathsf{RWD}$}
\newcommand{\crwd}{\mathsf{cRWD}}
\newcommand{\urwd}{\mathsf{uRWD}}
\newcommand{\g}{\mathsf{G}}
\newcommand{\ud}{\mathsf{D}^{\mathsf{U}}}
\newcommand{\cd}{\mathsf{D}^{\mathsf{C}}}
 
\usepackage{hyperref}
\usepackage{url}
\usepackage{todonotes}
\usepackage{subcaption}
\usepackage{yhmath}
\usepackage{booktabs}
\usepackage{multirow}

\newif\ifarxiv

\title{
High Fidelity Speech Synthesis \\
with Adversarial Networks
}

\author{Miko{\l}aj Bi\'nkowski\thanks{Work done at DeepMind.} \\
Department of Mathematics\\
Imperial College London\\
\texttt{mikbinkowski@gmail.com} \\
\AND
Jeff Donahue, Sander Dieleman, Aidan Clark, Erich Elsen, Norman Casagrande, \\
\textbf{Luis C. Cobo, Karen Simonyan} \\
DeepMind \\
\texttt{\{jeffdonahue,sedielem,aidanclark,eriche,ncasagrande,}\\
\texttt{luisca,simonyan\}@google.com} \\
}

\arxivtrue

\begin{document}

\maketitle

\begin{abstract}
    Generative adversarial networks have seen rapid development in recent years and have led to remarkable improvements in generative modelling of images. However, their application in the audio domain has received limited attention,
    and autoregressive models, such as WaveNet, remain the state of the art in generative modelling of audio signals such as human speech.
    To address this paucity, we introduce {\acronym}, a Generative Adversarial Network for Text-to-Speech.
    Our architecture is composed of a conditional feed-forward generator producing raw speech audio,
    and an ensemble of discriminators which operate on random windows of different sizes. The discriminators analyse the audio both in terms of general realism, as well as how well the audio corresponds to the utterance that should be pronounced. 
    To measure the performance of \acronym, we employ both subjective human evaluation (MOS -- Mean Opinion Score), as well as novel quantitative metrics (Fr\'echet DeepSpeech Distance and Kernel DeepSpeech Distance), which we find to be well correlated with MOS.
    We show that \acronym is capable of generating high-fidelity speech with naturalness comparable to the state-of-the-art models, and unlike autoregressive models, it is highly parallelisable thanks to an efficient feed-forward generator. Listen to \acronym reading this abstract at \url{https://storage.googleapis.com/deepmind-media/research/abstract.wav}.
\end{abstract}

\section{Introduction}
The Text-to-Speech (TTS) task consists in the conversion of text into speech audio. In recent years, the TTS field has seen remarkable progress, sparked by the development of neural autoregressive models for raw audio waveforms such as WaveNet~\citep{wavenet}, SampleRNN~\citep{samplernn} and WaveRNN~\citep{wavernn}. A notable limitation of these models is that they are difficult to parallelise over time: they predict each time step of an audio signal in sequence, which is computationally expensive and often impractical. A lot of recent research on neural models for TTS has focused on improving parallelism by predicting multiple time steps in parallel, e.g. using flow-based models~\citep{parallel-wavenet,clarinet,waveglow,flowavenet}. Such highly parallelisable models are more suitable to run efficiently on modern hardware.

An alternative approach for parallel waveform generation would be to use Generative Adversarial Networks~\citep[GANs,][]{gans}. GANs currently constitute one of the dominant paradigms for generative modelling of images, and they are able to produce high-fidelity samples that are almost indistinguishable from real data. However, their application to audio generation tasks has seen relatively limited success so far. In this paper, we explore raw waveform generation with GANs, and demonstrate that adversarially trained feed-forward generators are indeed able to synthesise high-fidelity speech audio. Our contributions are as follows:

\begin{itemize}
    \item We introduce \acronym, a Generative Adversarial Network for text-conditional high-fidelity speech synthesis. Its feed-forward generator is a convolutional neural network, coupled with an ensemble of multiple discriminators which evaluate the generated (and real) audio based on multi-frequency random windows. Notably, some discriminators take the linguistic conditioning into account (so they can measure how well the generated audio corresponds to the input utterance), while others ignore the conditioning, and can only assess the general realism of the audio.
    \item We propose a family of quantitative metrics for speech generation based on \emph{Fr\'echet Inception Distance}~\citep[FID,][]{fid} and \emph{Kernel Inception Distance}~\citep[KID,][]{kid}, where we replace the Inception image recognition network with the DeepSpeech audio recognition network.
    \item We present quantitative and subjective evaluation of TTS-GAN and its ablations, demonstrating the importance of our architectural choices. Our best-performing model achieves a MOS of $4.2$, which is comparable to the state-of-the-art WaveNet MOS of $4.4$, and establishes GANs as a viable option for efficient TTS.
\end{itemize}

\section{Related Work}

\subsection{Audio generation}
Most neural models for audio generation are likelihood-based: they represent an explicit probability distribution and the likelihood of the observed data is maximised under this distribution. Autoregressive models achieve this by factorising the joint distribution into a product of conditional distributions~\citep{wavenet,samplernn,wavernn,deepvoice}. Another strategy is to use an invertible feed-forward neural network to model the joint density directly~\citep{waveglow,flowavenet}. Alternatively, an invertible feed-forward model can be trained by distilling an autoregressive model using probability density distillation~\citep{parallel-wavenet,clarinet}, which enables it to focus on particular modes. Such mode-seeking behaviour is often desirable in conditional generation settings: we want the generated speech signals to sound realistic and correspond to the given text, but we are not interested in modelling every possible variation that occurs in the data. This reduces model capacity requirements, because parts of the data distribution may be ignored. Note that adversarial models exhibit similar behaviour, but without the distillation and invertibility requirements.

Many audio generation models, including all of those discussed so far, operate in the waveform domain: they directly model the amplitude of the waveform as it evolves over time. This is in stark contrast to most audio models designed for discriminative tasks (e.g. audio classification): 
such models tend to operate on time-frequency representations of audio (\emph{spectrograms}), which encode certain inductive biases with respect to the human perception of sound, 
and usually discard all phase information in the signal. While phase information is often inconsequential for discriminative tasks, generated audio signals must have a realistic phase component, because fidelity as judged by humans is severely affected otherwise. Because no special treatment for the phase component of the signal is required when generating directly in the waveform domain, this is usually more practical.

Tacotron~\citep{tacotron} and MelNet~\citep{melnet} constitute notable exceptions, and they use the Griffin-Lim algorithm~\citep{griffinlim} to reconstruct missing phase information, which the models themselves do not generate. Models like Deep Voice 2 \& 3~\citep{deepvoice2,deepvoice3} and Tacotron 2~\citep{tacotron2} achieve a compromise by first generating a spectral representation, and then using a separate autoregressive model to turn it into a waveform and fill in any missing spectral information. Because the generated spectrograms are imperfect, the waveform model has the additional task of correcting any mistakes. Char2wav~\citep{char2wav} uses intermediate vocoder features in a similar fashion.

\subsection{Generative adversarial networks}
Generative Adversarial Networks \citep[GANs,][]{gans} form a subclass of implicit generative models that relies on adversarial training of two networks: the \emph{generator}, which attempts to produce samples that mimic the reference distribution, and the \emph{discriminator}, which tries to differentiate between real and generated samples and, in doing so, provides a useful gradient signal to the generator.
Following rapid development, GANs have achieved state-of-the-art results in image~\citep{sagan,biggan,stylegan} and video generation~\citep{dvdgan}, and have been successfully applied for unsupervised feature learning~\citep{bigan, ali, bigbigan}, among many other applications.

Despite achieving impressive results in these domains, limited work has so far shown good performance of GANs in audio generation. Two notable exceptions include WaveGAN~\citep{wavegan} and GANSynth~\citep{gansynth}, which both successfully applied GANs to simple datasets of audio data. The former is the most similar to this work in the sense that it uses GANs to generate raw audio; results were obtained for a dataset of spoken commands of digits from zero to nine. The latter provides state-of-the-art results for a dataset of single note recordings from various musical instruments~\citep[NSynth, ][]{nsynth} by training GANs to generate invertible spectrograms of the notes. \citet{adversarial_vocoding} propose an adversarial vocoder model that is able to synthesise magnitude spectrograms from mel-spectrograms generated by Tacotron 2, and combine this with phase estimation using the Local Weighted Sums technique~\citep{lws}.

To the best of our knowledge, GANs have not yet been applied at large scale to non-visual domains. Two seconds of audio at 24kHz\footnote{24kHz is a commonly used frequency for speech, because the absence of frequencies above 12kHz does not meaningfully affect fidelity.} has a dimensionality of $48000$, which is comparable to RGB images at $128\times 128$ resolution. Until recently, high-quality GAN-generated images at such or higher resolution were uncommon~\citep{sagan,stylegan},
and it was not clear that training GANs at scale would lead to extensive improvements~\citep{biggan}.

Multiple discriminators have been used in GANs for different purposes. For images, \citet{lapgan,stackgan,progressive-growing} proposed to use separate discriminators for different resolutions. Similar approaches have also been used in image-to-image transfer \citep{munit} and video synthesis \citep{tganv2}. \citet{dvdgan}, on the other hand, combine a 3D-discriminator that scores the video at lower resolution and a 2D-frame discriminator which looks at individual frames. In adversarial feature learning, \citet{bigbigan} combine outputs from three discriminators to differentiate between joint distributions of images and latents. Discriminators operating on windows of the input have been used in adversarial texture synthesis~\citep{patchgan} and image translation~\citep{im2im,cyclegan}.

\section{\acronym}
\label{s:model}
\subsection{Dataset}
Our text-to-speech models are trained on a 
dataset which contains high-fidelity audio of human speech with the corresponding linguistic features and pitch information. The linguistic features encode phonetic and duration information, while the pitch is represented by the logarithmic fundamental frequency $\log F_0$. In total, there are $567$ features. We do not use ground-truth duration and pitch for subjective evaluation; we instead use duration and pitch predicted by separate models. 
The dataset is formed of variable-length audio clips containing single sequences, spoken by a professional voice actor in North American English. For training, we sample $2$ second windows (filtering out shorter examples) together with corresponding linguistic features. The total length of the filtered dataset is $44$ hours. The sampling frequency of the audio is $24$kHz, while the linguistic features and pitch are computed for $5$ms windows (at $200$Hz). This means that the generator network needs to learn how to convert the linguistic features and pitch into raw audio, while upsampling the signal by a factor of $120$. We apply a $\mu$-law transform to account for the logarithmic perception of volume (see Appendix \ref{a:mu-law}).

\subsection{Generator}

A summary of generator $\g$'s architecture is presented in Table \ref{t:generator} in Appendix \ref{a:architecture-details}.
The input to $\g$ is a sequence of linguistic and pitch features at $200$Hz, and its output is the raw waveform at $24$kHz.
The generator is composed of seven blocks (GBlocks, Figure~\ref{f:gblock}), each of which is a stack of two residual blocks~\citep{resnet_v2}.
As the generator is producing raw audio (e.g. a $2$s training clip corresponds to a sequence of $48000$ samples), we use dilated convolutions~\citep{dilated_conv} to ensure that the receptive field of $\g$ is large enough to capture long-term dependencies.
The four kernel size-$3$ convolutions in each GBlock have increasing dilation factors: $1,2,4,8$. Convolutions are preceded by Conditional Batch Normalisation~\citep{cond-batch-norm}, conditioned on the linear embeddings 
of the noise term $z\sim\mathcal{N}(0,\mI_{128})$ in the single-speaker case, or the concatenation of $z$ and a one-hot representation of the speaker ID in the multi-speaker case. The embeddings are different for each BatchNorm instance.
A GBlock contains two skip connections, the first of which performs upsampling if the output frequency is higher than the input, and it also contains a size-$1$ convolution if the number of output channels is different from the input.
GBlocks 3--7 gradually upsample the temporal dimension of hidden representations by factors of $2,2,2,3,5$, while the number of channels is reduced by GBlocks 3, 6 and 7 (by a factor of $2$ each).
The final convolutional layer with \emph{Tanh} activation produces a single-channel audio waveform.

\subsection{Ensemble of Random Window Discriminators}
\label{s:rwds}
Instead of a single discriminator, we use an ensemble of Random Window Discriminators ({\rwd}s) which operate on randomly sub-sampled fragments of the real or generated samples. 
The ensemble allows for the evaluation of audio in different complementary ways, and is obtained by taking a Cartesian product of two parameter spaces: 
(i) the size of the random windows fed into the discriminator;
(ii) whether a discriminator is conditioned on linguistic and pitch features.
For example, in our best-performing model, we consider five window sizes ($240,480,960,1920,3600$ samples), which yields $10$ discriminators in total. Notably, the number of discriminators only affects the training computation requirements, as at inference time only the generator network is used, while the discriminators are discarded. However, thanks to the use of relatively short random windows, the proposed ensemble leads to faster training than conventional discriminators.

Using random windows of different size, rather than the full generated sample, has a data augmentation effect and also reduces the computational complexity of {\rwd}s, as explained next.
In the first layer of each discriminator, we reshape (downsample) the input raw waveform to a constant temporal dimension $\omega=240$ by moving consecutive blocks of samples into the channel dimension, i.e.\ from $[\omega k,1]$ to $[\omega,k]$, where $k$ is the downsampling factor (e.g.\ $k=8$ for input window size $1920$). This way, all the {\rwd}s have the same architecture and similar computational complexity despite different window sizes.
We confirm these design choices experimentally in Section~\ref{s:experiments}.

The conditional discriminators have access to linguistic and pitch features, and can measure whether the generated audio matches the input conditioning. This means that random windows in conditional discriminators need to be aligned with the conditioning frequency to preserve the correspondence between the waveform and linguistic features within the sampled window. This limits the valid sampling to that of the frequency of the conditioning signal ($200$Hz, or every $5$ms).
The unconditional discriminators, on the contrary, only evaluate whether the generated audio sounds realistic regardless of the conditioning. The random windows for these discriminators are sampled without constraints at full $24$kHz frequency, which further increases the amount of training data.
More formally, we define conditional and unconditional {\rwd}s as stochastic functions:
\begin{align}
    \crwd_{k, \omega}(x, c; \theta) &= \cd_k(x_{\rj:\rj + \omega k}, c_{\rj/\lambda: (\rj + \omega k)/\lambda}; \theta),\qquad \rj\sim\mathcal{U}\left(\left\{0,\lambda, 2\lambda,\ldots,N - \omega k\right\}\right) \\
    \urwd_{k, \omega}(x; \theta) &= \ud_k(x_{\rj:\rj + \omega k}; \theta),\qquad \rj\sim\mathcal{U}\left(\left\{0,1,\ldots,N -\omega k\right\}\right),
\end{align}
where $x$ and $c$ are respectively the waveform and linguistic features, $\theta$ is set of network parameters, and $\lambda=120$ is a frequency ratio between $x$ and $c$.

The final ensemble discriminator combines $10$ different {\rwd}'s:
\begin{equation}
    \mathsf{RWD}^*_{\omega}(x, c; \theta^*) = \sum_{k\in\{1,2,4,8,15\}}\crwd_{k, \omega}(x, c; \theta_k) + \urwd_{k, \omega}(x; \theta_k'), \qquad \theta^* = \bigcup_k(\theta_k \cup \theta_k').
\end{equation}
In Section \ref{s:experiments} we describe other combinations of {\rwd}s as well as a full, deterministic discriminator which we used in our ablation study.

\subsection{Discriminator Architecture}
\begin{figure}[ht]
    \centering
    \vspace*{-15pt}
    \scalebox{0.9}{
    \begin{subfigure}[b]{0.54\textwidth}
        \centering
        \includegraphics[width=.9\textwidth]{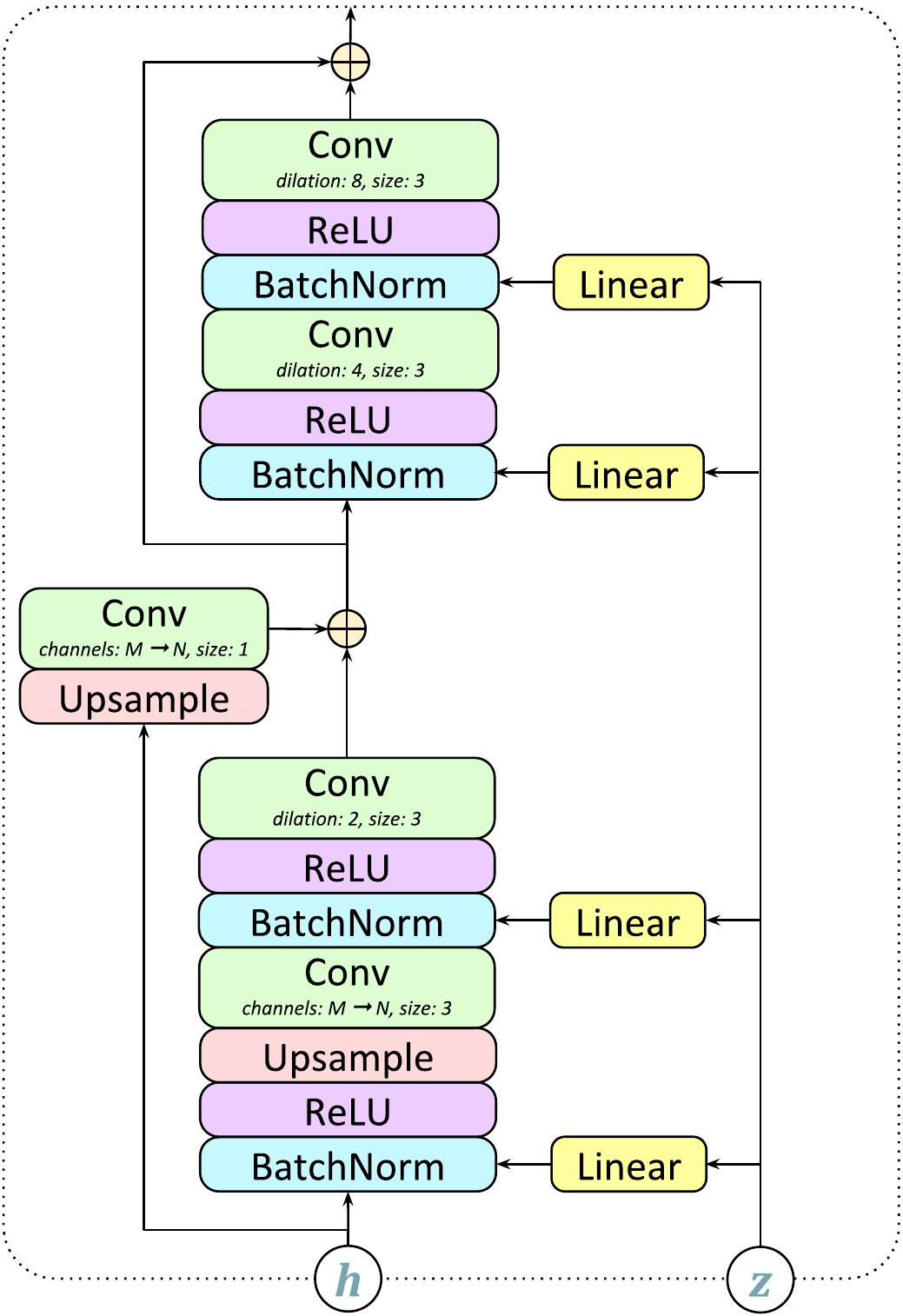}
        \caption{GBlock}
        \label{f:gblock}
    \end{subfigure}
    \begin{subfigure}[b]{0.45\textwidth}
        \centering
        \begin{subfigure}[b]{\textwidth}
            \centering
            \includegraphics[width=.9\textwidth]{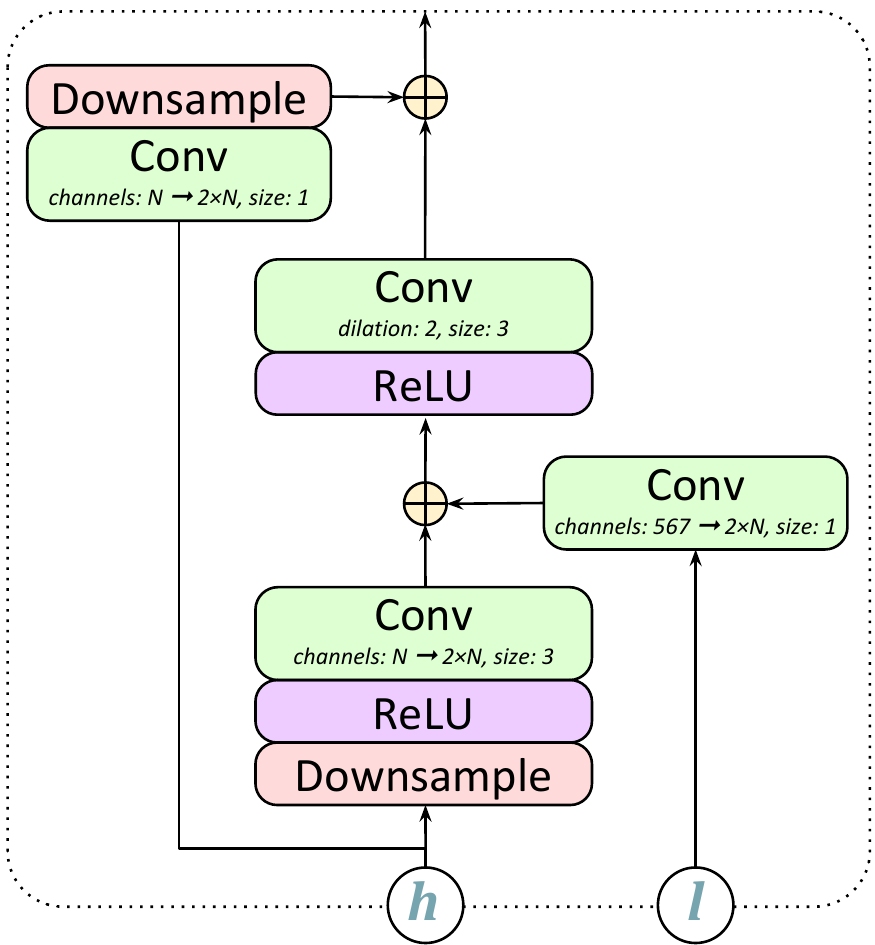}
            \caption{Conditional DBlock}
            \label{f:cdblock}
        \end{subfigure}
        \begin{subfigure}[b]{\textwidth}
            \centering
            \includegraphics[width=.8\textwidth]{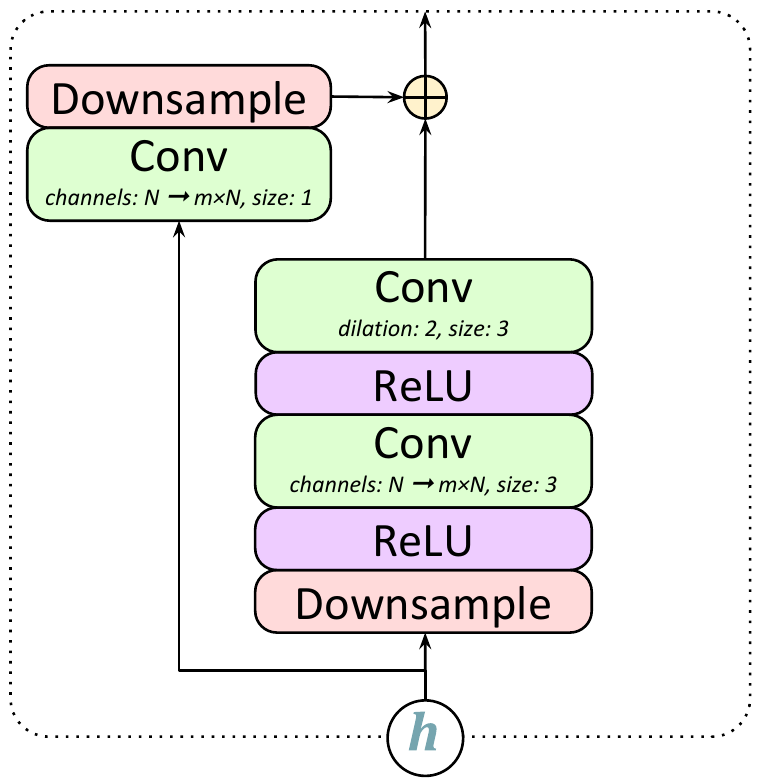}
            \caption{DBlock}
            \label{f:dblock}
        \end{subfigure}
    \end{subfigure}
    }
    \caption{Residual blocks used in the model. Convolutional layers have the same number of input and output channels and no dilation unless stated otherwise. $h$ - hidden layer representation, $l$ - linguistic features, $z$ - noise vector, $m\in\{1, 2\}$ - channel multiplier, $M$- \textsf{G}'s input channels, $M=2N$ in blocks 3, 6, 7, and $M=N$ otherwise; \emph{size} refers to kernel size.}
    \label{f:blocks}
\end{figure}
\begin{figure}[ht]
    \centering
    \includegraphics[width=.9\textwidth]{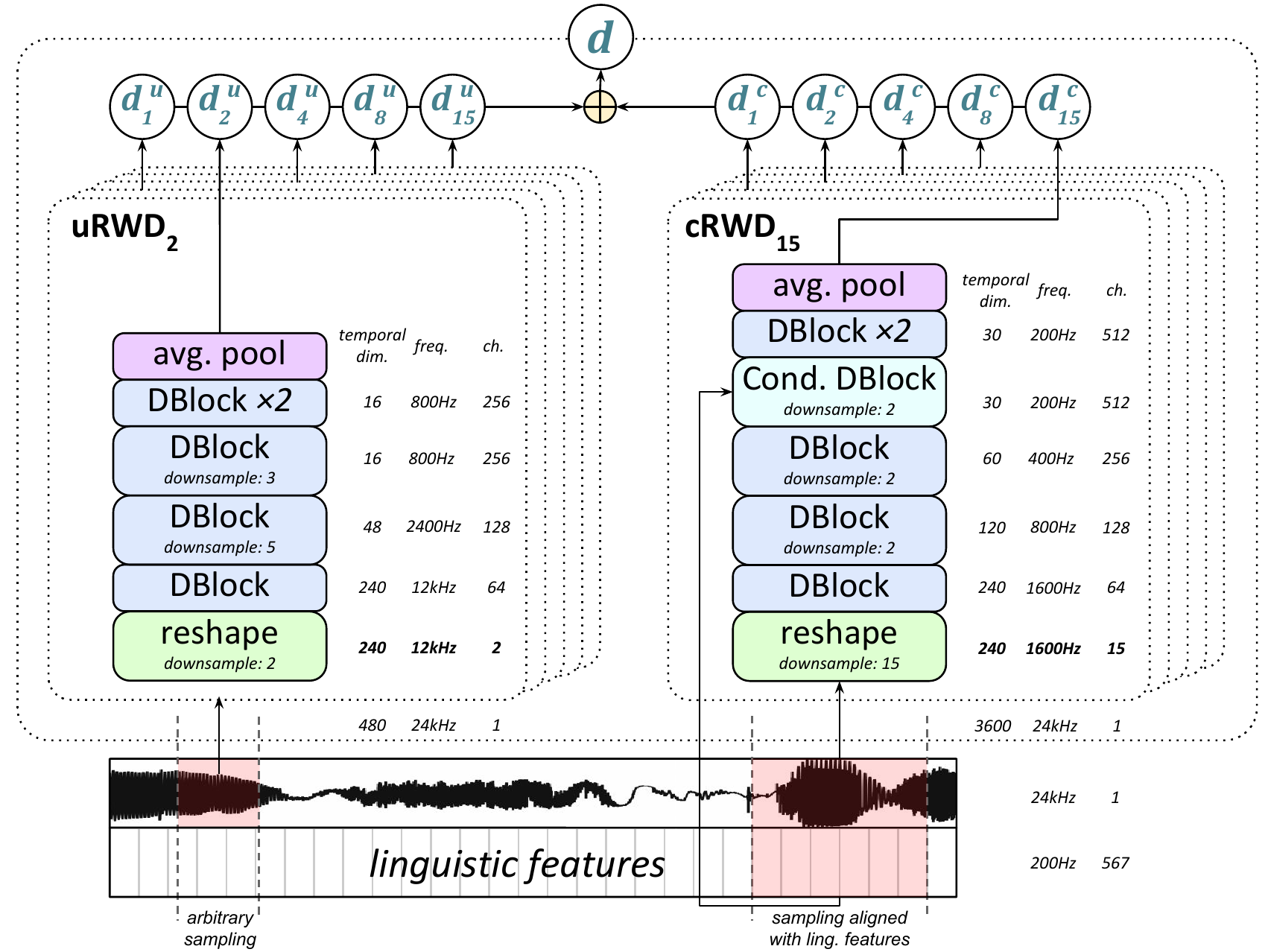}
    \caption{\emph{Multiple Random Window Discriminator} architecture. The discriminator combines outputs from 5 unconditional ($\urwd$s, left) and 5 conditional ($\crwd$s, right) discriminators; one of each group is detailed in the diagram. The number of downsampling blocks is fixed for $\urwd$s and depends on the input window size $\omega k$ for $\crwd$s, see Table \ref{t:downsample-factors}. $ch.$ - number of output channels.}
    \label{f:discriminator}
\end{figure}

The full discriminator architecture is shown in Figure~\ref{f:discriminator}. The discriminators consists of blocks (DBlocks) that are similar to the GBlocks used in the generator, but without batch normalisation. The architectures of standard and conditional DBlocks are shown in Figures~\ref{f:cdblock} and ~\ref{f:dblock} respectively. The only difference between the two DBlocks is that in the conditional DBlock, the embedding of the linguistic features is added after the first convolution.
The first and the last two DBlocks do not downsample (i.e.\ keep the temporal dimension fixed). Apart from that, we add at least two downsampling blocks in the middle, with downsample factors depending on $k$, so as to match the frequency of the linguistic features (see Appendix~\ref{a:architecture-details} for details).
Unconditional {\rwd}s are composed entirely of DBlocks.
In conditional {\rwd}s, the input waveform is gradually downsampled by DBlocks, until the temporal dimension of the activation is equal to that of the conditioning, at which point a conditional DBlock is used. This joint information is then passed to the remaining DBlocks, whose final output is average-pooled to obtain a scalar.
The dilation factors in the DBlocks' convolutions follow the pattern $1,2,1,2$ -- unlike the generator, the discriminator operates on a relatively small window, and we did not observe any benefit from using larger dilation factors.

\section{Evaluation}
We provide subjective human evaluation of our model using Mean Opinion Scores (\emph{MOS}), as well as quantitative metrics.

\subsection{MOS}
We evaluate our model on a set of $1000$ sentences, using human evaluators. Each evaluator was asked to mark the subjective \emph{naturalness} of a sentence on a 1-5 Likert scale, comparing to the scores reported by \citet{parallel-wavenet} for WaveNet and Parallel WaveNet.

Although our model was trained to generate 2 second audio clips with the starting point not necessarily aligned with the beginning of a sentence, we are able to generate samples of arbitrary length. This is feasible due to the fully convolutional nature of the generator and carried out using a \emph{convolutional masking} trick, detailed in Appendix \ref{a:longer-samples}. Human evaluators scored full sentences with a length of up to 15 seconds.

\subsection{Speech Distances}
We introduce a family of quantitative metrics for generative models of speech, which include the unconditional and conditional \emph{Fr\'echet DeepSpeech Distance (\textsf{FDSD, cFDSD})} and \emph{Kernel DeepSpeech Distance (\textsf{KDSD, cKDSD})}. These metrics follow common metrics used in evaluation of GANs for images, \emph{Fr\'echet Inception Distance}~\citep[FID,][]{fid} and \emph{Kernel Inception Distance}~\citep[KID,][]{kid}.

FID and KID compute the \emph{Fr\'echet distance} and the \emph{Maximum Mean Discrepancy}~\citep[MMD,][]{mmd} respectively between representations of reference and generated distributions extracted from a pre-trained \emph{Inception} network~\citep{inception_v3}. To obtain analogous metrics for speech, we extract the features from an open-source implementation of an accurate speech recognition model, \emph{DeepSpeech2}~\citep{deepspeech2}. Specifically, we use the implementation available in the \emph{NVIDIA OpenSeq2Seq} library~\citep{openseq2seq} and extract features from the last layer, whose output is used in the CTC loss during training. We use representations in the resulting feature space to compute the Fr\'echet distance and MMD (See Appendix \ref{a:deepspeech2} for details). 

We note that \citet{fad} proposed a similar metric, \emph{Fr\'echet Audio Distance}. This metric, however, has been designed for music datasets and uses a music classifier as a feature extractor; therefore it is not well-suited to evaluate text-to-speech models.

As conditioning plays a crucial role in our task, we compute two variants of these metrics, conditional (\textsf{cFDSD, cKDSD}) and unconditional (\textsf{FDSD, KDSD}). Both Fr\'echet and Kernel distance provide scores with respect to a \emph{reference real} sample and require both the real sample and the generated one to be independent and identically distributed.
Assume that variables $\rx^{real}$ and $\rx^{\g}$ are drawn from the real and and generated distributions, while $\rc$ is drawn from the distribution of linguistic features. In the conditional case, \textsf{cFDSD} and \textsf{cKDSD} compute distances between conditional distributions $p(\rx^{G}|\rc)$ and $p(\rx^{real}|\rc)$. In the unconditional case, \textsf{FDSD} and \textsf{KDSD} compare $p(\rx^{G})$ and $p(\rx^{real})$.

Both metrics are estimated using 10,000 generated and reference samples, drawn independently with the same (in the conditional case), or independent (in the unconditional case) linguistic features. This procedure is detailed in Appendix \ref{a:distance_estimation}.

The main reason for using both Fr\'echet and Kernel distances is the popularity of FID in the image domain, despite the issue of its biased estimator, as shown by \citet{kid}. Thanks to the availability of an unbiased estimator of MMD, this issue does not apply to kernel-based distances. For instance, they yield zero values for real data, which allows comparison in the conditional case. We give more details on these distances in Appendix \ref{a:metrics}.

\section{Experiments}\label{s:experiments}
In this section we discuss the experiments, comparing \acronym with WaveNet and carrying out ablations that validate our architectural choices.

As mentioned in Section \ref{s:model}, the main architectural choices made in our model include the use of multiple {\rwd}s, conditional and unconditional, with a number of different downsampling factors. 
We thus consider the following ablations of our best-performing model:
\begin{enumerate}
    \item full-input discriminator $\mathsf{FullD} = \cd_1$,
    \item single conditional {\rwd}: $\crwd_1$,
    \item multiple conditional {\rwd}s: $\crwd_{\{1,2,4,8,15\}} = \sum_{k\in\{1,2,4,8,15\}}\crwd_k$,    
    \item single conditional and single unconditional {\rwd}: $\crwd_1 + \urwd_1$,
    \item five independent $\crwd$s and $\urwd$s: \newline$\left(\crwd_1 + \urwd_1\right)^{\times5}(x, c) := \sum_{i=1}^5\crwd_1(x,c; \theta_i) + \urwd_1(x; \theta'_i)$,
    \item 10 {\rwd}s without downsampling but with different window sizes:\newline
    $
    \mathsf{RWD}_{1, 240\times\{1,2,4,8,15\}} = \sum_{k\in\{1,2,4,8,15\}}\left(\crwd_{1, 240k} + \urwd_{1, 240k}\right)$
    \item 10 {\rwd}s with longer window: $\mathsf{RWD}^*_{480}$.
\end{enumerate}
All other parameters of these models were the same as in the proposed one. In Appendix \ref{a:training-details} we present details of the hyperparameters used during training.

\subsection{Results}
\begin{table}[h]
    \centering
    \resizebox{\columnwidth}{!}{
        \begin{tabular}{lccccc}
             model & MOS & $\mathsf{FDSD}$ & $\mathsf{cFDSD}$  & $\begin{array}{c} \mathsf{KDSD} \\ \times 10^{5}\end{array}$ & $\begin{array}{c} \mathsf{cKDSD} \\ \times 10^{5}\end{array}$ \\ \hline\hline
             \emph{natural speech} & $4.55 \pm 0.075$ & $0.161$ & N/A & $0$ & $0$ \\ \hline
             \emph{WaveNet}, \citet{wavenet} &  $4.41 \pm 0.069$ \\
             \emph{Parallel WaveNet}, \citet{parallel-wavenet} &  $4.41 \pm 0.078$ \\ \hline
             
             $\mathsf{FullD}$ & $1.889 \pm 0.057$ & $4.51$ & $4.46$ & $785$ & $782$ \\ 
             
             $\crwd_1$ & $3.394 \pm 0.058$ & $0.362$ & $0.247$ & $35.2$ & $30.9$ \\ 
             
             $\crwd_{\{1,2,4,8,15\}}$ & $3.498 \pm 0.059$ & $0.398$ & $0.284$ & $42.1$ & $37.9$ \\ 
             
             $\crwd_1+\urwd_1$ & $3.502 \pm 0.057$ & $0.259$ & $0.144$ & $16.6$ & $12.3$ \\
             
             $\left(\crwd_1 + \urwd_1\right)^{\times5}$ & $3.526 \pm 0.054$ & $0.194$ & $0.073$ & $5.59$ & $1.34$ \\

             $\mathsf{RWD}_{1, 240\times\{1,2,4,8,15\}}$ & $4.154 \pm 0.050$ & $0.184$ & $0.061$ & $3.73$ & $0.54$ \\ 
             
             $\mathsf{RWD}^*_{480}$ & $4.195 \pm 0.045$ & $0.193$ & $0.069$ & $5.28$ & $0.98$ \\ \hline 
             
             \acronym ($\mathsf{RWD}^*$) & $4.213 \pm 0.046$ & $0.184$ & $0.060$ & $3.84$ & $0.37$ \\ \hline 
        \end{tabular}
    }
    \caption{Results from prior work, the ablation study and the proposed model. Mean opinion scores for natural speech, WaveNet and Parallel WaveNet are taken from \citet{parallel-wavenet} and are not directly comparable due to dataset differences. For natural speech we present estimated \textsf{FDSD} -- non-zero due to the bias of the estimator -- and theoretical values of \textsf{KDSD} and \textsf{cKDSD}. \textsf{cFDSD} is unavailable; see Appendix \ref{a:metrics}.
    }
    \label{t:results}
\end{table}
Table \ref{t:results} presents quantitative evaluations of the proposed model, together with benchmarks and other variants of {\acronym} that we considered in this work. 

Our best model achieves worse yet comparable scores to the strong baselines, WaveNet and Parallel WaveNet. This performance, however, has not yet been achieved using adversarial techniques and is still very good, especially when compared to parametric text-to-speech models. These results are however not directly comparable due to dataset differences; for instance WaveNet and Parallel WaveNet were trained on 65 hours of data, a bit more than \acronym.

Our ablation study confirms the importance of combining multiple {\rwd}s. The deterministic full discriminator achieved the worst scores. All multiple-{\rwd} models achieved better results than a single $\crwd_1$; all models that used unconditional {\rwd}s were better than those that did not. Comparing 10-discriminator models, it is clear that combinations of different window sizes were beneficial, as a simple ensemble of 10 fixed-size windows was significantly worse. All three 10-{\rwd} models with varying discriminator sizes achieved similar mean opinion scores, with the downsampling model with base window size 240 performing best.

We also observe a noticeable correlation between human evaluation scores (MOS) and the proposed metrics, which demonstrates that these metrics are well-suited for the evaluation of neural audio synthesis models.

\subsection{Discussion}
\paragraph{Random window discriminators.}
Although it is difficult to say why {\rwd}s work much better than the full discriminator, we conjecture that this is because of the relative simplicity of the distributions that the former must discriminate between, and the number of different samples we can draw from these distributions. For example, the largest window discriminators used in our best model discriminate between distributions supported on $\R^{3600}$, and there are respectively 371 and 44,401 different windows that can be sub-sampled from a 2s clip (real or generated) by conditional and unconditional {\rwd}s of effective window size 3600. The full discriminator, on the other hand, always sees full real or generated examples sampled from a distribution supported on $\R^{48000}$.

\paragraph{Computational efficiency.}
Our Generator has a larger receptive field (590ms, i.e. 118 steps at the frequency of the linguistic features) and three times fewer FLOPs (0.64 MFLOP/sample) than Parallel WaveNet (receptive field size: 320ms, 1.97 MFLOP/sample). However, the discriminators used in our ensemble compare windows of shorter sizes, from 10ms to 150ms. Since these windows are much shorter than the entire generated clips, training with ensembles of such {\rwd}s is faster than with $\mathsf{FullD}$. In terms of depth, our generator has 30 layers, which is a half of Parallel WaveNet's, while the depths of the discriminators vary between 11 and 17 layers, as discussed in Appendix \ref{a:architecture-details}.

\paragraph{Stability.}
The proposed model enjoyed very stable training, with gradual improvement of subjective sample quality and decreasing values of the proposed metrics. Despite training for as many as 1 million steps, we have not experienced model collapses often reported in GAN literature and studied in detail by \citet{biggan}.

\section{Conclusion}
We have introduced \acronym, a GAN for raw audio text-to-speech generation. 
Unlike state-of-the-art text-to-speech models, \acronym is adversarially trained and the resulting generator is a feed-forward convolutional network. This allows for very efficient audio generation, which is important in practical applications.
Our architectural exploration lead to the development of a model with an ensemble of unconditional and conditional \emph{Random Window Discriminators} operating at different window sizes, which respectively assess the realism of the generated speech and its correspondence with the input text.
We showed in an ablation study that each of these components is instrumental to achieving good performance.
We have also proposed a family of quantitative metrics for generative models of speech: \emph{(conditional) Fr\'echet DeepSpeech Distance} and \emph{(conditional) Kernel DeepSpeech Distance}, and demonstrated experimentally that these metrics rank models in line with Mean Opinion Scores obtained through human evaluation.
As they are based on the publicly available DeepSpeech recognition model, they will be made available for the machine learning community.
Our quantitative results as well as subjective evaluation of the generated samples showcase the feasibility of text-to-speech generation with GANs.

\subsubsection*{Acknowledgments}
We would like to thank A{\"a}ron van den Oord, Andrew Brock and the rest of the DeepMind team.

\bibliography{ttsgan}

\begin{thebibliography}{52}
\providecommand{\natexlab}[1]{#1}
\providecommand{\url}[1]{\texttt{#1}}
\expandafter\ifx\csname urlstyle\endcsname\relax
  \providecommand{\doi}[1]{doi: #1}\else
  \providecommand{\doi}{doi: \begingroup \urlstyle{rm}\Url}\fi

\bibitem[Amodei et~al.(2016)Amodei, Anubhai, Battenberg, Case, Casper,
  Catanzaro, Chen, Chrzanowski, Coates, Diamos, Elsen, Engel, Fan, Fougner,
  Han, Hannun, Jun, LeGresley, Lin, Narang, Ng, Ozair, Prenger, Raiman,
  Satheesh, Seetapun, Sengupta, Wang, Wang, Wang, Xiao, Yogatama, Zhan, and
  Zhu]{deepspeech2}
Dario Amodei, Rishita Anubhai, Eric Battenberg, Carl Case, Jared Casper, Bryan
  Catanzaro, Jingdong Chen, Mike Chrzanowski, Adam Coates, Greg Diamos, Erich
  Elsen, Jesse Engel, Linxi Fan, Christopher Fougner, Tony Han, Awni Hannun,
  Billy Jun, Patrick LeGresley, Libby Lin, Sharan Narang, Andrew Ng, Sherjil
  Ozair, Ryan Prenger, Jonathan Raiman, Sanjeev Satheesh, David Seetapun,
  Shubho Sengupta, Yi~Wang, Zhiqian Wang, Chong Wang, Bo~Xiao, Dani Yogatama,
  Jun Zhan, and Zhenyao Zhu.
\newblock Deep {S}peech 2: End-to-end speech recognition in {E}nglish and
  {M}andarin.
\newblock In \emph{ICML}, 2016.

\bibitem[Arik et~al.(2017)Arik, Chrzanowski, Coates, Diamos, Gibiansky, Kang,
  Li, Miller, Ng, Raiman, et~al.]{deepvoice}
Sercan~{\"O} Arik, Mike Chrzanowski, Adam Coates, Gregory Diamos, Andrew
  Gibiansky, Yongguo Kang, Xian Li, John Miller, Andrew Ng, Jonathan Raiman,
  et~al.
\newblock Deep {V}oice: Real-time neural text-to-speech.
\newblock In \emph{ICML}, 2017.

\bibitem[Bi{\'n}kowski et~al.(2018)Bi{\'n}kowski, Sutherland, Arbel, and
  Gretton]{kid}
Miko{\l}aj Bi{\'n}kowski, Dougal~J Sutherland, Michael Arbel, and Arthur
  Gretton.
\newblock Demystifying {MMD} {GAN}s.
\newblock In \emph{ICLR}, 2018.

\bibitem[Brock et~al.(2016)Brock, Lim, Ritchie, and Weston]{ortho-reg}
Andrew Brock, Theodore Lim, James~M. Ritchie, and Nick Weston.
\newblock Neural photo editing with introspective adversarial networks.
\newblock In \emph{ICLR}, 2016.

\bibitem[Brock et~al.(2019)Brock, Donahue, and Simonyan]{biggan}
Andrew Brock, Jeff Donahue, and Karen Simonyan.
\newblock Large scale {GAN} training for high fidelity natural image synthesis.
\newblock In \emph{ICLR}, 2019.

\bibitem[Clark et~al.(2019)Clark, Donahue, and Simonyan]{dvdgan}
Aidan Clark, Jeff Donahue, and Karen Simonyan.
\newblock Efficient video generation on complex datasets.
\newblock \emph{arXiv:1907.06571}, 2019.

\bibitem[Denton et~al.(2015)Denton, Chintala, Szlam, and Fergus]{lapgan}
Emily~L Denton, Soumith Chintala, Arthur Szlam, and Rob Fergus.
\newblock Deep generative image models using a {L}aplacian pyramid of
  adversarial networks.
\newblock In \emph{NeurIPS}, 2015.

\bibitem[Donahue et~al.(2019)Donahue, McAuley, and Puckette]{wavegan}
Chris Donahue, Julian McAuley, and Miller Puckette.
\newblock Adversarial audio synthesis.
\newblock In \emph{ICLR}, 2019.

\bibitem[Donahue \& Simonyan(2019)Donahue and Simonyan]{bigbigan}
Jeff Donahue and Karen Simonyan.
\newblock Large scale adversarial representation learning.
\newblock \emph{arXiv:1907.02544}, 2019.

\bibitem[Donahue et~al.(2017)Donahue, Kr{\"{a}}henb{\"{u}}hl, and
  Darrell]{bigan}
Jeff Donahue, Philipp Kr{\"{a}}henb{\"{u}}hl, and Trevor Darrell.
\newblock Adversarial feature learning.
\newblock In \emph{ICLR}, 2017.

\bibitem[Dumoulin et~al.(2017{\natexlab{a}})Dumoulin, Belghazi, Poole,
  Mastropietro, Lamb, Arjovsky, and Courville]{ali}
Vincent Dumoulin, Ishmael Belghazi, Ben Poole, Olivier Mastropietro, Alex Lamb,
  Martin Arjovsky, and Aaron Courville.
\newblock Adversarially learned inference.
\newblock In \emph{ICLR}, 2017{\natexlab{a}}.

\bibitem[Dumoulin et~al.(2017{\natexlab{b}})Dumoulin, Shlens, and
  Kudlur]{cond-batch-norm}
Vincent Dumoulin, Jonathon Shlens, and Manjunath Kudlur.
\newblock A learned representation for artistic style.
\newblock In \emph{ICLR}, 2017{\natexlab{b}}.

\bibitem[Engel et~al.(2017)Engel, Resnick, Roberts, Dieleman, Norouzi, Eck, and
  Simonyan]{nsynth}
Jesse Engel, Cinjon Resnick, Adam Roberts, Sander Dieleman, Mohammad Norouzi,
  Douglas Eck, and Karen Simonyan.
\newblock Neural audio synthesis of musical notes with {WaveNet} autoencoders.
\newblock In \emph{ICML}, 2017.

\bibitem[Engel et~al.(2019)Engel, Agrawal, Chen, Gulrajani, Donahue, and
  Roberts]{gansynth}
Jesse Engel, Kumar~Krishna Agrawal, Shuo Chen, Ishaan Gulrajani, Chris Donahue,
  and Adam Roberts.
\newblock {GANS}ynth: Adversarial neural audio synthesis.
\newblock In \emph{ICLR}, 2019.

\bibitem[Gibiansky et~al.(2017)Gibiansky, Arik, Diamos, Miller, Peng, Ping,
  Raiman, and Zhou]{deepvoice2}
Andrew Gibiansky, Sercan Arik, Gregory Diamos, John Miller, Kainan Peng, Wei
  Ping, Jonathan Raiman, and Yanqi Zhou.
\newblock Deep {V}oice 2: Multi-speaker neural text-to-speech.
\newblock In \emph{NeurIPS}, 2017.

\bibitem[Goodfellow et~al.(2014)Goodfellow, Pouget-Abadie, Mirza, Xu,
  Warde-Farley, Ozair, Courville, and Bengio]{gans}
Ian Goodfellow, Jean Pouget-Abadie, Mehdi Mirza, Bing Xu, David Warde-Farley,
  Sherjil Ozair, Aaron Courville, and Yoshua Bengio.
\newblock Generative adversarial nets.
\newblock In \emph{NeurIPS}, 2014.

\bibitem[Gretton et~al.(2012)Gretton, Borgwardt, Rasch, Sch{\"o}lkopf, and
  Smola]{mmd}
A.~Gretton, K.~Borgwardt, M.~Rasch, B.~Sch{\"o}lkopf, and A.~Smola.
\newblock A kernel two-sample test.
\newblock \emph{JMLR}, 2012.

\bibitem[Griffin \& Lim(1984)Griffin and Lim]{griffinlim}
Daniel Griffin and Jae Lim.
\newblock Signal estimation from modified short-time {F}ourier transform.
\newblock \emph{IEEE Transactions on Acoustics, Speech, and Signal Processing},
  1984.

\bibitem[He et~al.(2016)He, Zhang, Ren, and Sun]{resnet_v2}
Kaiming He, Xiangyu Zhang, Shaoqing Ren, and Jian Sun.
\newblock Identity mappings in deep residual networks.
\newblock In \emph{ECCV}, 2016.

\bibitem[Heusel et~al.(2017)Heusel, Ramsauer, Unterthiner, Nessler, and
  Hochreiter]{fid}
Martin Heusel, Hubert Ramsauer, Thomas Unterthiner, Bernhard Nessler, and Sepp
  Hochreiter.
\newblock {GAN}s trained by a two time-scale update rule converge to a local
  {N}ash equilibrium.
\newblock In \emph{NeurIPS}, 2017.

\bibitem[Huang et~al.(2018)Huang, Liu, Belongie, and Kautz]{munit}
Xun Huang, Ming-Yu Liu, Serge Belongie, and Jan Kautz.
\newblock Multimodal unsupervised image-to-image translation.
\newblock In \emph{ECCV}, 2018.

\bibitem[Ioffe \& Szegedy(2015)Ioffe and Szegedy]{batch-norm}
Sergey Ioffe and Christian Szegedy.
\newblock Batch normalization: Accelerating deep network training by reducing
  internal covariate shift.
\newblock In \emph{ICML}, 2015.

\bibitem[Isola et~al.(2017)Isola, Zhu, Zhou, and Efros]{im2im}
Phillip Isola, Jun-Yan Zhu, Tinghui Zhou, and Alexei~A Efros.
\newblock Image-to-image translation with conditional adversarial networks.
\newblock In \emph{CVPR}, 2017.

\bibitem[Kalchbrenner et~al.(2018)Kalchbrenner, Elsen, Simonyan, Noury,
  Casagrande, Lockhart, Stimberg, van~den Oord, Dieleman, and
  Kavukcuoglu]{wavernn}
Nal Kalchbrenner, Erich Elsen, Karen Simonyan, Seb Noury, Norman Casagrande,
  Edward Lockhart, Florian Stimberg, A{\"{a}}ron van~den Oord, Sander Dieleman,
  and Koray Kavukcuoglu.
\newblock Efficient neural audio synthesis.
\newblock In \emph{ICML}, 2018.

\bibitem[Karras et~al.(2018)Karras, Aila, Laine, and
  Lehtinen]{progressive-growing}
Tero Karras, Timo Aila, Samuli Laine, and Jaakko Lehtinen.
\newblock Progressive growing of {GAN}s for improved quality, stability, and
  variation.
\newblock In \emph{ICLR}, 2018.

\bibitem[Karras et~al.(2019)Karras, Laine, and Aila]{stylegan}
Tero Karras, Samuli Laine, and Timo Aila.
\newblock A style-based generator architecture for generative adversarial
  networks.
\newblock In \emph{CVPR}, 2019.

\bibitem[Kilgour et~al.(2019)Kilgour, Zuluaga, Roblek, and Sharifi]{fad}
Kevin Kilgour, Mauricio Zuluaga, Dominik Roblek, and Matthew Sharifi.
\newblock Fr\'echet audio distance: A metric for evaluating music enhancement
  algorithms.
\newblock In \emph{Interspeech}, 2019.

\bibitem[Kim et~al.(2019)Kim, Lee, Song, Kim, and Yoon]{flowavenet}
Sungwon Kim, Sang-Gil Lee, Jongyoon Song, Jaehyeon Kim, and Sungroh Yoon.
\newblock Flo{W}ave{N}et: A generative flow for raw audio.
\newblock In \emph{ICML}, 2019.

\bibitem[Kingma \& Ba(2015)Kingma and Ba]{adam}
Diederik~P Kingma and Jimmy Ba.
\newblock Adam: A method for stochastic optimization.
\newblock In \emph{ICLR}, 2015.

\bibitem[Kuchaiev et~al.(2018)Kuchaiev, Ginsburg, Gitman, Lavrukhin, Case, and
  Micikevicius]{openseq2seq}
Oleksii Kuchaiev, Boris Ginsburg, Igor Gitman, Vitaly Lavrukhin, Carl Case, and
  Paulius Micikevicius.
\newblock {O}pen{S}eq2{S}eq: Extensible toolkit for distributed and mixed
  precision training of sequence-to-sequence models.
\newblock In \emph{NLP-OSS}, 2018.

\bibitem[{Le Roux} et~al.(2010){Le Roux}, Kameoka, Ono, and Sagayama]{lws}
Jonathan {Le Roux}, Hirokazu Kameoka, Nobutaka Ono, and Shigeki Sagayama.
\newblock Fast signal reconstruction from magnitude {STFT} spectrogram based on
  spectrogram consistency.
\newblock In \emph{DAFx}, 2010.

\bibitem[Li \& Wand(2016)Li and Wand]{patchgan}
Chuan Li and Michael Wand.
\newblock Precomputed real-time texture synthesis with {M}arkovian generative
  adversarial networks.
\newblock In \emph{ECCV}, 2016.

\bibitem[Lim \& Ye(2017)Lim and Ye]{geometricgan}
Jae~Hyun Lim and Jong~Chul Ye.
\newblock Geometric {GAN}.
\newblock \emph{arXiv:1705.02894}, 2017.

\bibitem[Mehri et~al.(2017)Mehri, Kumar, Gulrajani, Kumar, Jain, Sotelo,
  Courville, and Bengio]{samplernn}
Soroush Mehri, Kundan Kumar, Ishaan Gulrajani, Rithesh Kumar, Shubham Jain,
  Jose Sotelo, Aaron Courville, and Yoshua Bengio.
\newblock {SampleRNN}: An unconditional end-to-end neural audio generation
  model.
\newblock In \emph{ICLR}, 2017.

\bibitem[Miyato et~al.(2018)Miyato, Kataoka, Koyama, and Yoshida]{spectralnorm}
Takeru Miyato, Toshiki Kataoka, Masanori Koyama, and Yuichi Yoshida.
\newblock Spectral normalization for generative adversarial networks.
\newblock In \emph{ICLR}, 2018.

\bibitem[Neekhara et~al.(2019)Neekhara, Donahue, Puckette, Dubnov, and
  McAuley]{adversarial_vocoding}
Paarth Neekhara, Chris Donahue, Miller Puckette, Shlomo Dubnov, and Julian
  McAuley.
\newblock Expediting {TTS} synthesis with adversarial vocoding.
\newblock In \emph{Interspeech}, 2019.

\bibitem[Ping et~al.(2018)Ping, Peng, Gibiansky, Arik, Kannan, Narang, Raiman,
  and Miller]{deepvoice3}
Wei Ping, Kainan Peng, Andrew Gibiansky, Sercan~O. Arik, Ajay Kannan, Sharan
  Narang, Jonathan Raiman, and John Miller.
\newblock Deep {V}oice 3: 2000-speaker neural text-to-speech.
\newblock In \emph{ICLR}, 2018.

\bibitem[Ping et~al.(2019)Ping, Peng, and Chen]{clarinet}
Wei Ping, Kainan Peng, and Jitong Chen.
\newblock Clari{N}et: Parallel wave generation in end-to-end text-to-speech.
\newblock In \emph{ICLR}, 2019.

\bibitem[Prenger et~al.(2019)Prenger, Valle, and Catanzaro]{waveglow}
Ryan Prenger, Rafael Valle, and Bryan Catanzaro.
\newblock Wave{G}low: A flow-based generative network for speech synthesis.
\newblock In \emph{ICASSP}, 2019.

\bibitem[Saito \& Saito(2018)Saito and Saito]{tganv2}
Masaki Saito and Shunta Saito.
\newblock {TGANv2}: Efficient training of large models for video generation
  with multiple subsampling layers.
\newblock \emph{arXiv:1811.09245}, 2018.

\bibitem[Saxe et~al.(2014)Saxe, McClelland, and Ganguli]{ortho-init}
Andrew Saxe, James McClelland, and Surya Ganguli.
\newblock Exact solutions to the nonlinear dynamics of learning in deep linear
  neural networks.
\newblock In \emph{ICLR}, 2014.

\bibitem[Shen et~al.(2018)Shen, Pang, Weiss, Schuster, Jaitly, Yang, Chen,
  Zhang, Wang, Skerrv-Ryan, et~al.]{tacotron2}
Jonathan Shen, Ruoming Pang, Ron~J Weiss, Mike Schuster, Navdeep Jaitly,
  Zongheng Yang, Zhifeng Chen, Yu~Zhang, Yuxuan Wang, Rj~Skerrv-Ryan, et~al.
\newblock Natural {TTS} synthesis by conditioning {W}ave{N}et on {M}el
  spectrogram predictions.
\newblock In \emph{ICASSP}, 2018.

\bibitem[Sotelo et~al.(2017)Sotelo, Mehri, Kumar, Santos, Kastner, Courville,
  and Bengio]{char2wav}
Jose Sotelo, Soroush Mehri, Kundan Kumar, Joao~Felipe Santos, Kyle Kastner,
  Aaron Courville, and Yoshua Bengio.
\newblock Char2{W}av: End-to-end speech synthesis.
\newblock In \emph{ICLR}, 2017.

\bibitem[Szegedy et~al.(2016)Szegedy, Vanhoucke, Ioffe, Shlens, and
  Wojna]{inception_v3}
Christian Szegedy, Vincent Vanhoucke, Sergey Ioffe, Jon Shlens, and Zbigniew
  Wojna.
\newblock Rethinking the {I}nception architecture for computer vision.
\newblock In \emph{CVPR}, 2016.

\bibitem[van~den Oord et~al.(2016)van~den Oord, Dieleman, Zen, Simonyan,
  Vinyals, Graves, Kalchbrenner, Senior, and Kavukcuoglu]{wavenet}
A{\"{a}}ron van~den Oord, Sander Dieleman, Heiga Zen, Karen Simonyan, Oriol
  Vinyals, Alex Graves, Nal Kalchbrenner, Andrew~W. Senior, and Koray
  Kavukcuoglu.
\newblock Wave{N}et: A generative model for raw audio.
\newblock \emph{arXiv:1609.03499}, 2016.

\bibitem[van~den Oord et~al.(2018)van~den Oord, Li, Babuschkin, Simonyan,
  Vinyals, Kavukcuoglu, Driessche, Lockhart, Cobo, Stimberg, Casagrande, Grewe,
  Noury, Dieleman, Elsen, Kalchbrenner, Zen, Graves, King, Walters, Belov, and
  Hassabis]{parallel-wavenet}
A{\"{a}}ron van~den Oord, Yazhe Li, Igor Babuschkin, Karen Simonyan, Oriol
  Vinyals, Koray Kavukcuoglu, George Driessche, Edward Lockhart, Luis Cobo,
  Florian Stimberg, Norman Casagrande, Dominik Grewe, Seb Noury, Sander
  Dieleman, Erich Elsen, Nal Kalchbrenner, Heiga Zen, Alex Graves, Helen King,
  Tom Walters, Dan Belov, and Demis Hassabis.
\newblock Parallel {WaveNet}: Fast high-fidelity speech synthesis.
\newblock In \emph{ICML}, 2018.

\bibitem[Vasquez \& Lewis(2019)Vasquez and Lewis]{melnet}
Sean Vasquez and Mike Lewis.
\newblock Mel{N}et: A generative model for audio in the frequency domain.
\newblock \emph{arXiv:1906.01083}, 2019.

\bibitem[Wang et~al.(2017)Wang, Skerry-Ryan, Stanton, Wu, Weiss, Jaitly, Yang,
  Xiao, Chen, Bengio, et~al.]{tacotron}
Yuxuan Wang, RJ~Skerry-Ryan, Daisy Stanton, Yonghui Wu, Ron~J Weiss, Navdeep
  Jaitly, Zongheng Yang, Ying Xiao, Zhifeng Chen, Samy Bengio, et~al.
\newblock Tacotron: Towards end-to-end speech synthesis.
\newblock In \emph{Interspeech}, 2017.

\bibitem[Yu \& Koltun(2016)Yu and Koltun]{dilated_conv}
Fisher Yu and Vladlen Koltun.
\newblock Multi-scale context aggregation by dilated convolutions.
\newblock In \emph{ICLR}, 2016.

\bibitem[Zhang et~al.(2017)Zhang, Xu, Li, Zhang, Wang, Huang, and
  Metaxas]{stackgan}
Han Zhang, Tao Xu, Hongsheng Li, Shaoting Zhang, Xiaogang Wang, Xiaolei Huang,
  and Dimitris~N Metaxas.
\newblock Stackgan: Text to photo-realistic image synthesis with stacked
  generative adversarial networks.
\newblock In \emph{ICCV}, 2017.

\bibitem[Zhang et~al.(2019)Zhang, Goodfellow, Metaxas, and Odena]{sagan}
Han Zhang, Ian Goodfellow, Dimitris Metaxas, and Augustus Odena.
\newblock Self-attention generative adversarial networks.
\newblock In \emph{ICML}, 2019.

\bibitem[Zhu et~al.(2017)Zhu, Park, Isola, and Efros]{cyclegan}
Jun-Yan Zhu, Taesung Park, Phillip Isola, and Alexei~A Efros.
\newblock Unpaired image-to-image translation using cycle-consistent
  adversarial networks.
\newblock In \emph{ICCV}, 2017.

\end{thebibliography}
\bibliographystyle{iclr2020_conference}

\newpage
\appendix
\section{Architecture details}
\subsection{Masking convolutions to generate longer samples} \label{a:longer-samples}
\begin{figure}
    \centering
    \includegraphics[width=\textwidth]{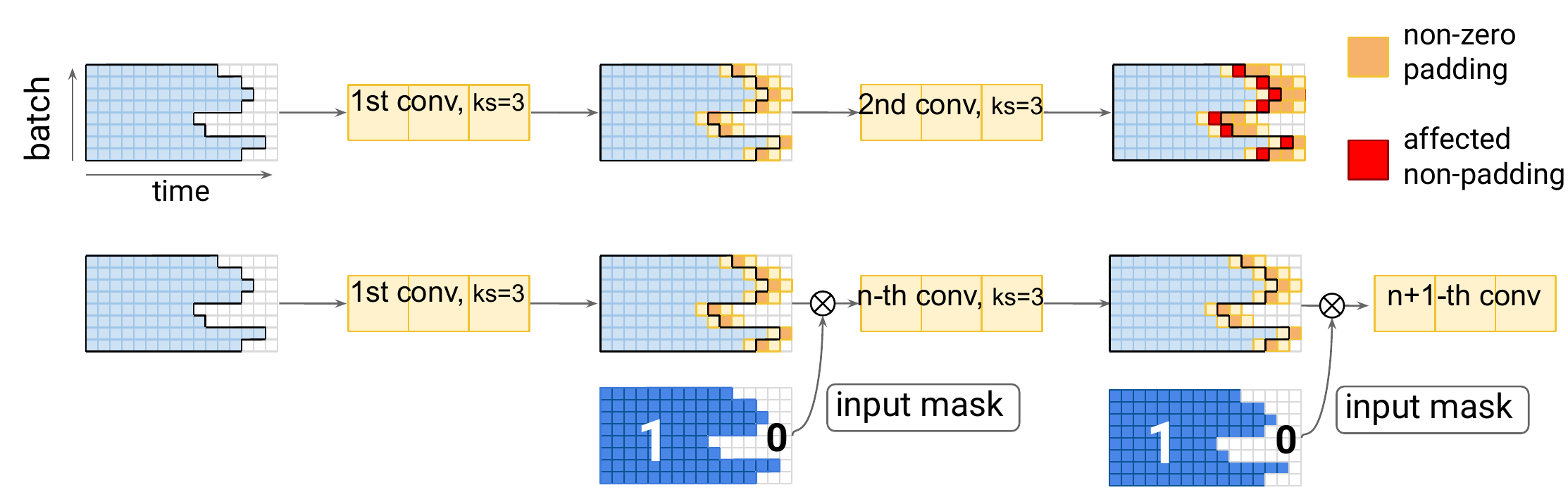}
    \caption{Masking scheme for sampling different-length samples. Top: processing a batch of samples of different lengths padded with zeros leads to interference between padding and non-padding after the second convolution, not seen during training. Bottom: masking after each convolution ensures that the meaningful input seen by each layer is padded with zeros only.}
    \label{f:masking}
\end{figure}
Since our generator is a fully-convolutional network, in theory it is capable of generating samples of arbitrary length. However, since deep learning frameworks usually require processing fixed-size samples in batches for efficiency reasons, our inputs of different lengths need to be zero-padded to fit in a fixed-size tensor. Convolutional layers, including the ones used in our model, often pad their inputs to create outputs of the desired dimensionality, hence we only need to ensure that the padded part of the input tensors to all layers is always zero. As shown in Figure \ref{f:masking}, this would not normally be the case after the second convolutional layer, since convolutions (with kernel sizes greater than one) would propagate non-zero values outside the border between meaningful input and padding. A simple way to address this issue is masking, i.e. multiplying the input by a zero-one mask tensor, directly before each convolutional layer. This enables batched sampling of utterances of different length, which is efficient on many hardware platforms, optimised for batching.

\subsection{Architecture details}
\label{a:architecture-details}
\begin{table}[ht]
    \centering
    \begin{tabular}{l|ccc}
        layer/input & $t$ & \emph{freq.} & $ch$ \\
        \cmidrule(lr){1-1}\cmidrule(lr){2-4}
        \cmidrule(lr){1-1}\cmidrule(lr){2-4}
        \emph{linguistic features}, $z$ & 400 & 200Hz & 567 \\
        \cmidrule(lr){1-1}\cmidrule(lr){2-4}
        conv, \emph{kernel size 3} & 400 & 200Hz & 768 \\
        GBlock & 400 & 200Hz & 768 \\
        GBlock & 400 & 200Hz & 768 \\
        GBlock, \emph{upsample} $\times 2$ & 800 & 400Hz & 384 \\
        GBlock, \emph{upsample} $\times 2$ & 1600 & 800Hz & 384 \\
        GBlock, \emph{upsample} $\times 2$ & 3200 & 1600Hz & 384 \\
        GBlock, \emph{upsample} $\times 3$ & 9600 & 4800Hz & 192 \\
        GBlock, \emph{upsample} $\times 5$ & 48000 & 24kHz & 96 \\
        conv, \emph{kernel size 3} & 48000 & 24kHz & 1 \\
        Tanh & & & \\
        \cmidrule(lr){1-1}\cmidrule(lr){2-4}
        \cmidrule(lr){1-1}\cmidrule(lr){2-4}        
    \end{tabular}
    \caption{Architecture of \acronym's Generator. $t$ denotes the temporal dimension, while $ch$ denotes the number of channels. The rightmost three columns describe dimensions of the \emph{output} of the corresponding layer.}
    \label{t:generator}
\end{table}
\begin{table}
    \centering
    \begin{tabular}{ccccccc}
         $k$ & \multicolumn{3}{c}{$\cd_k$} & \multicolumn{3}{c}{$\ud_k$} \\
         \cmidrule(lr){2-4}\cmidrule(lr){5-7}
         {}  & \emph{factors} & num. blocks & depth  & \emph{factors} & num. blocks & depth\\
         \cmidrule(lr){1-7}
         $1$ & $5,3,2,2,2$ & $8$ & $17$ &  $5,3$ & $5$ & $11$\\
         $2$ & $5,3,2,2$ & $7$ & $15$ & $5,3$ & $5$ & $11$ \\
         $4$ & $5,3,2$ & $6$ & $13$ & $5,3$ & $5$ & $11$ \\
         $8$ & $5,3$ & $5$ & $11$ & $5,3$ & $5$ & $11$\\
         $15$ & $2,2,2$ & $6$  & $13$ & $2,2$ & $5$ & $11$ \\
    \end{tabular}
    \caption{Downsample factors in discriminators for different initial stride values $k$. The number of blocks includes non-downsampling DBlocks.}
    \label{t:downsample-factors}
\end{table}
In Table \ref{t:generator} we present the details of Generator architecture. Overall, the generator has 30 layers, most of which are parts of dilated residual blocks.

Table \ref{t:downsample-factors} shows the numbers of residual DBlocks and downsample factors in these blocks for different initial downsample factors of {\rwd}s.

All conditional discriminators eventually add the representations of the waveform and the linguistic features. This happens once the temporal dimension of the main residual stack is downsampled to the dimension of the linguistic features, i.e. by a factor of 120. Downsampling is carried out via an initial reshape operation (by a factor $k$ varying per {\rwd}) and then in residual blocks, whose downsample factors are prime divisors of $120/k$, in decreasing order. For unconditional discriminators, we use only the first two largest prime divisors of $120/k$.

\section{DeepSpeech distances - details}
\label{a:speech_distances}

\subsection{DeepSpeech2}
\label{a:deepspeech2}
Our evaluation metrics extract high-level features from raw audio using the pre-trained DeepSpeech2 model from the \emph{NVIDIA OpenSeq2Seq} library~\citep{openseq2seq}. Let $w=480$ be a $20$ms window of raw audio at 24kHz, and let $f:\R^{w}\longrightarrow\R^{1600}$ be a function that maps such a window through the DeepSpeech2 network up to the 1600-dimensional output of the layer labeled \verb+ForwardPass/ds2_encoder/Reshape_2:0+. We use default values for all settings of the DeepSpeech2 model; $f$ also includes the model's preprocessing layers.

For a 2s audio clip $\va\in\R^{100w}$, we define
\begin{equation} \label{eq:ds}
    \mathsf{DS}(\va) = \frac{1}{199}\sum_{i=0}^{198} f(\va_{iw/2:iw/2 + w})\in\R^{1600},
\end{equation}
where $\va_{i:j} = (\va_i, \va_{i+1}, \ldots, \va_{j-1})'$ is a vector slice.

The function $\mathsf{DS}$ therefore computes 1600 features for each 20ms window, sampled evenly with 10ms overlap, and then takes the average of the features along the temporal dimension. 

\subsection{Metrics in distribution space}
\label{a:metrics}
Given samples $\mX\in\R^{m\times d}$ and $\mY\in\R^{n\times d}$, where $d$ is the representation dimension, the Fr\'echet distance and MMD can be computed using the following estimators:
\begin{align}
    \widehat{\textrm{Fr\'echet}^2}(\mX, \mY) =& \|\mu_{\mX} - \mu_{\mY}\|_2^2 + \mathrm{Tr}\left(\Sigma_{\mX} + \Sigma_{\mY} - 2 (\Sigma_{\mX}\Sigma_{\mY})^{1/2}\right) \label{eq:frechet}\\
    \widehat{\textrm{MMD}^2}(\mX, \mY) =& \frac{1}{m(m-1)}\sum_{\substack{1\leq i,j\leq m \\ i\neq j}} k(\mX_i, \mX_j) + \frac{1}{n(n-1)}\sum_{\substack{1\leq i,j\leq n \\ i\neq j}} k(\mY_i, \mY_j) \nonumber \\
    & + \sum_{i=1}^m\sum_{j=1}^n k(\mX_i, \mY_j), \label{eq:mmd}
\end{align}
where $\mu_{\mX}, \mu_{\mY}$ and $\Sigma_{\mX}, \Sigma_{\mY}$ are the means and covariance matrices of $\mX$ and $\mY$ respectively, while $k: \R^d\times\R^d\longrightarrow\R$ is a positive definite kernel function. Following \cite{kid} we use the polynomial kernel
\begin{equation}
    k(x,y) = \left(\tfrac{1}{d}x^T y + 1\right)^3.
\end{equation}

Estimator (\ref{eq:frechet}) has been found to be biased~\citep{kid}, even for large sample sizes. For this reason, FID estimates for real data (i.e. when $\mX$ and $\mY$ are both drawn independently from the same distribution) are positive, even though the theoretical value of such a metric is zero. KID, however, does not suffer from this issue thanks to the use of the unbiased estimator (\ref{eq:mmd}). These properties also apply to the proposed DeepSpeech metrics.

The lack of bias in an estimator is particularly important for establishing scores on real data for conditional distances. In our conditional text-to-speech setting, we cannot sample two independent real samples with the same conditioning, and for this reason we cannot estimate the value of \textsf{cFDSD} for real data, which would be positive due to bias of estimator (\ref{eq:frechet}). For \textsf{cKDSD}, however, we know that such an estimator would have given values very close to zero, if we had been able to evaluate it on two real i.i.d. samples with the same conditioning.

\subsection{Distance estimation} \label{a:distance_estimation}
Let $\mathsf{G}$ and $\mathsf{DS}$ represent the generator function and a function that maps audio to DeepSpeech2 features as defined in Eq. \ref{eq:ds}. Let 
\begin{equation}
    \mX^{\g} = \{\mathsf{DS}\left(\g(c_i, z_i)\right)\}_{i=1}^N, \qquad \mX^{real}_{:N} = \{\mathsf{DS}(x_i)\}_{i=1}^{N}, \qquad \mX^{real}_{N:} = \{\mathsf{DS}(x_i)\}_{i=N+1}^{2N},
\end{equation}
where $(x_i, c_i) \overset{iid}{\sim} p(\rx^{real}, \rc), i=1,\ldots,2N$ are jointly sampled real examples and linguistic features, and $z_i\overset{iid}{\sim}\mathcal{N}(0, 1)$. 
In the conditional case, we use the same conditioning in the reference and generated samples, comparing conditional distributions $p(\rx^{G}|\rc)$ and $p(\rx^{real}|\rc)$:
\begin{align}
    \widehat{\mathsf{cFDSD}}\left(p(\rx^{\g}|\rc), p(\rx^{real}|\rc)\right) &= \widehat{\textrm{Fr\'echet}}\left(\mX^{\g}, \mX^{real}_{:N}\right), \\
    \widehat{\mathsf{cKDSD}}\left(p(\rx^G|\rc), p(\rx^{real}|\rc)\right) &= \widehat{\textrm{MMD}}\left(\mX^{\g}, \mX^{real}_{:N}\right),
\end{align}
where $\widehat{\textrm{Fr\'echet}}$ and $\widehat{\textrm{MMD}}$ are estimators of the Fr\'echet distance and MMD defined in Eq. \ref{eq:frechet} and \ref{eq:mmd}, respectively.

In the unconditional case, we compare $p(\rx^{G})$ and $p(\rx^{real})$: 
\begin{align}
    \widehat{\mathsf{FDSD}}\left(p(\rx^{\g}), p(\rx^{real})\right) &= \widehat{\textrm{Fr\'echet}}\left(\mX^{\g}, \mX^{real}_{N:}\right), \\ 
    \widehat{\mathsf{KDSD}}\left(p(\rx^{\g}), p(\rx^{real})\right) &= \widehat{\textrm{MMD}}\left(\mX^{\g}, \mX^{real}_{N:}\right).
\end{align}

\section{$\mu$-law preprocessing} \label{a:mu-law}
Many generative models of audio use the $\mu$-law transform to account for the logarithmic perception of volume. Although $\mu$-law is typically used in the context of non-uniform quantisation, we use the transform without the quantisation step as our model operates in the continuous domain:
\begin{equation}
    F(x) = \textrm{sgn}(x)\frac{\ln(1 + \mu|x|)}{\ln(1 + \mu)},
\end{equation}
where $x\in[-1, 1]$ and $\mu=2^8-1=255$ for 8-bit encoding or $\mu=2^{16}-1=65,535$ for 16-bit encoding.

Our early experiments showed better performance of models generating $\mu$-law transformed audio than non-transformed waveforms. We used the 16-bit transformation.

\section{Training details} \label{a:training-details}
\begin{figure}
    \centering
    \includegraphics[width=.9\textwidth]{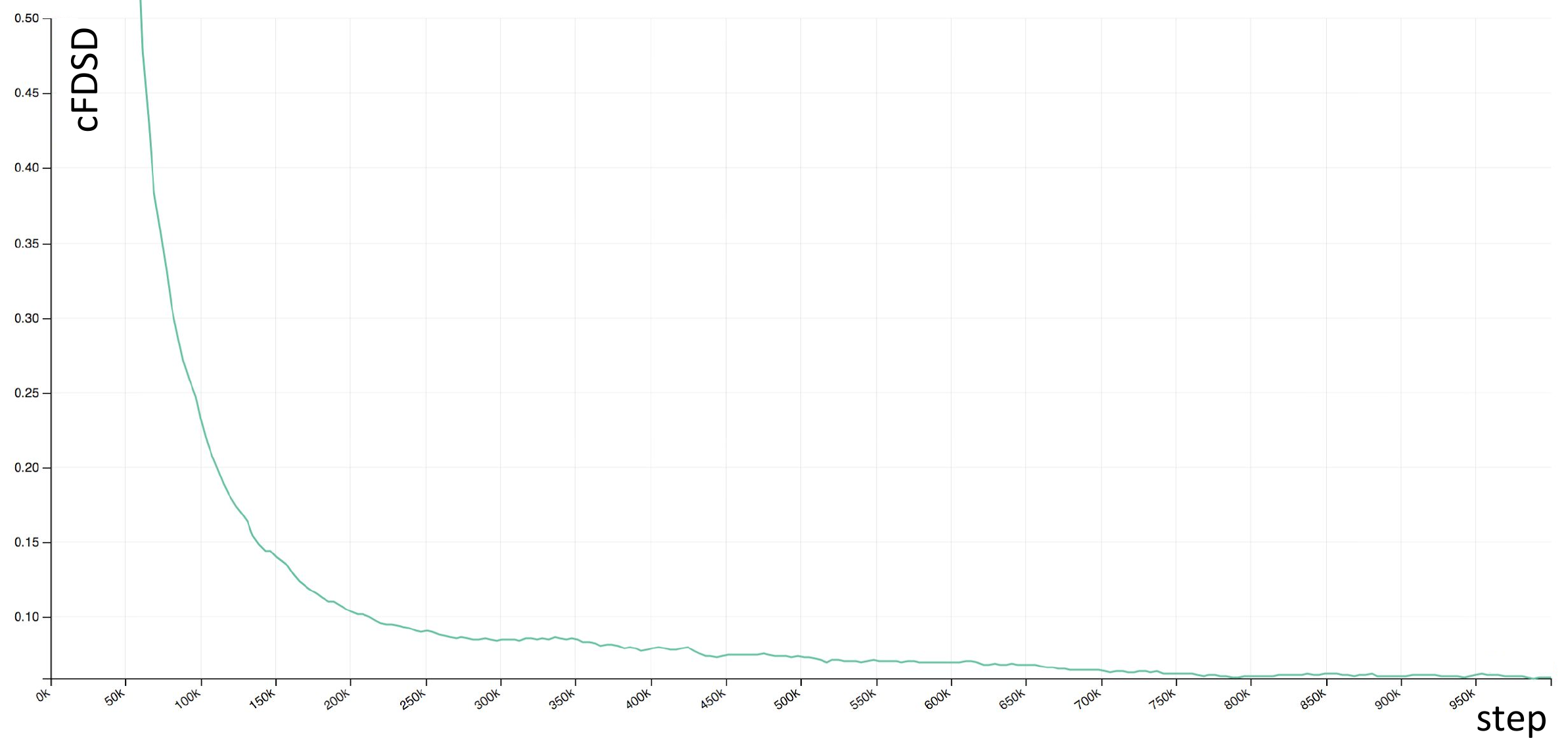}
    \caption{Learning curve for the \acronym model in terms of \textsf{cFDSD}.}
    \label{f:learning_curves}
\end{figure}
We train all models with a single discriminator step per generator step, but with doubled learning rate: $10^{-4}$ for the former, compared to $5\times10^{-5}$ for the latter. We use the \emph{hinge loss}~\citep{geometricgan}, a batch size of $1024$ and the Adam optimizer~\citep{adam} with hyperparameters $\beta_1=0, \beta_2=0.999$.

Following \citet{biggan}, we use spectral normalisation \citep{spectralnorm} and orthogonal initialisation \citep{ortho-init} in both the generator and discriminator(s), and apply off-diagonal orthogonal regularisation \citep{ortho-reg,biggan} and exponential moving averaging to the generator weights with a decay rate of 0.9999 for sampling. We also use \emph{cross-replica} BatchNorm \citep{batch-norm}, which aggregates batch statistics from all devices across which the batch is split and \emph{standing statistics} during sampling.
The latter means that we accumulate batch statistics from 100 forward passes through the generator before the actual sampling takes place, allowing for inference at arbitrary batch sizes.

In fact, accumulating standing statistics makes the BatchNorm layers in the generator independent of any characteristics of the samples produced during inference.
This technique is thus vital for sampling audio of unspecified length: producing samples that are longer than those used during training typically requires using a smaller batch size, with partially padded samples (See Appendix \ref{a:longer-samples}). These smaller batches would naturally have different statistics than the batches used during training. 

We trained our models on Cloud TPU v3 Pods with data parallelism over 128 replicas for 1 million generator and discriminator updates, which usually took up to 48 hours.

Figure \ref{f:learning_curves} presents the stable and gradual decrease of \textsf{cFDSD} during training.

\end{document}